\documentclass[reqno,11pt]{amsart}
\usepackage{graphicx}
\usepackage[mathscr]{eucal}
\usepackage{amssymb}
\numberwithin{equation}{section}

\title[A Spatially Homogeneous and Isotropic Einstein-Dirac Cosmology]{A Spatially Homogeneous and Isotropic Einstein-Dirac Cosmology}

\author[F.\ Finster]{Felix Finster}
\address{Fakult\"at f\"ur Mathematik, Universit\"at Regensburg, D-93040 Regensburg, Germany}
\email{Felix.Finster@mathematik.uni-regensburg.de}
\thanks{F.F.\ is partially supported by the Deutsche Forschungsgemeinschaft. }

\author[C.\ Hainzl]{Christian Hainzl \\ \\ January 2011}
\address{Mathematisches Institut, Universit\"at T\"ubingen,  
D-72076 T\"ubingen, Germany}
\email{christian.hainzl@uni-tuebingen.de}
\thanks{C.H.\ is partially supported by U.S. National Science
Foundation, grant DMS-0800906.}

\newtheorem{Def}{Definition}[section]
\newtheorem{Thm}[Def]{Theorem}
\newtheorem{Prp}[Def]{Proposition}
\newtheorem{Lemma}[Def]{Lemma}

\newcommand{\Thanks}{\vspace*{.5em} \noindent \thanks}
\newcommand{\beq}{\begin{equation}}
\newcommand{\eeq}{\end{equation}}
\newcommand{\Proof}{\begin{proof}}
\newcommand{\QED}{\end{proof} \noindent}

\newcommand{\Rmax}{R_\text{\rm{max}}}
\newcommand{\Rtilt}{R_\text{\rm{tilt}}}
\newcommand{\Rqu}{R_\text{\rm{qu}}}
\newcommand{\wmax}{w_{1,{\rm max}}}
\newcommand{\ttilt}{t_{\rm tilt}}

\newcommand{\C}{\mathbb{C}}
\newcommand{\R}{\mathbb{R}}
\newcommand{\1}{\mbox{\rm 1 \hspace{-1.05 em} 1}}
\newcommand{\Z}{\mathbb{Z}}
\newcommand{\N}{\mathbb{N}}
\newcommand{\D}{\mathcal{D}}
\newcommand{\eps}{\varepsilon}

\DeclareMathOperator{\Tr}{Tr}
\DeclareMathOperator{\tr}{tr}

\allowdisplaybreaks[1]

\begin{document}
\maketitle

\begin{abstract}
We consider a spatially homogeneous and isotropic cosmological model where Dirac spinors
are coupled to classical gravity. For the Dirac spinors we choose a Hartree-Fock ansatz
where all one-particle wave functions are coherent
and have the same momentum. If the scale function is large, the universe behaves like the
classical Friedmann dust solution. If however the scale function is small, quantum effects
lead to oscillations of the energy-momentum tensor.
It is shown numerically and proven analytically that these quantum oscillations
can prevent the formation of a big bang or big crunch singularity. 
The energy conditions are analyzed. We prove the existence
of time-periodic solutions which go through an infinite number of expansion and contraction
cycles.
\end{abstract}

\tableofcontents

\newpage
\section{Introduction}
The nonlinear coupling of gravity as described by Einstein's equations to matter formed of
quantum mechanical waves has led to surprising physical effects and has incited the development
of interesting mathematical methods needed for their analysis. T.D.\ Lee et al~\cite{lee}
studied a complex massive scalar field and found spherically symmetric, static soliton solutions,
interpreted physically as ``mini-soliton stars.''
In the seminal papers~\cite{christodoulou1, christodoulou2, christodoulou3},
D.\ Christodoulou analyzed the spherically symmetric gravitational collapse of a real massless scalar field.
The study of more fundamental quantum mechanical matter formed of Dirac wave functions
was initiated in~\cite{singlet}, where spherically symmetric, static soliton solutions of the coupled
Einstein-Dirac (ED) equations were constructed numerically. Moreover, it was shown that the
Einstein-Dirac-Maxwell and Einstein-Dirac-Yang/Mills equations do not admit spherically
symmetric, static black-hole solutions (see~\cite[Part~II]{survey} and the references therein
as well as~\cite{yannPhD, yann}). The resulting physical picture for the gravitational collapse is that
after a horizon has formed, the Dirac particles necessarily either disappear in the black hole
or escape to infinity, but they cannot remain in a bounded region outside the horizon.
More recently, in the independent works~\cite{bird, stuart, rota} existence results were proved for
the spherically symmetric, static ED solitons, using quite different mathematical methods.

For time-dependent, spherically symmetric ED systems, the numerical results of~\cite{ventrella} as
well as the improved and more detailed numerical analysis in~\cite{zeller+hiptmair, zeller} show
that a ``cloud'' of Dirac particles either dissipates to infinity or else a black hole forms, in which case
the Dirac wave partially disappears inside the horizon and escapes to infinity.
Unfortunately, apart from the dynamical stability analysis in~\cite{singlet},
there are no rigorous results on time-dependent, spherically symmetric ED systems.
The main problem is that, compared to the massless scalar field in~\cite{christodoulou1},
the rest mass and the fact that the Dirac spinors have several components cause additional
difficulties. This was our main motivation for searching for a simpler time-dependent ED system
which is easier to analyze but nevertheless reveals new effects on the nonlinear dynamics.
This led us to considering spatially homogeneous and isotropic systems. By separating the
spatial dependence, one can reduce the system to nonlinear ordinary differential equations in time.
In the recent paper~\cite{friedmann}, it was shown numerically that 
in such systems, quantum oscillations of the Dirac wave functions can prevent the
formation of a big bang or big crunch singularity. In the present work, we give a detailed
derivation of the underlying equations and present a rigorous analysis of all quantum effects.

More specifically, we consider the simplest possible cosmological model where the metric is given by the
homogeneous and isotropic line element
\beq\label{lineel} ds^2 = dt^2 - R^2(t) \left( \frac {dr^2}{1 - k r^2}  + r^2 d\Omega^2 \right) , \eeq
where~$d\Omega^2$ is the line element on~$S^2$, and~$R(t)$ is the so-called {\em{scale function}}.
The parameter~$k$ can take the values~$-1$, $0$ or~$1$, corresponding to an open, a flat and a closed
universe, respectively. We will mainly focus on the closed case, but we will give a brief
outlook on the other cases as well. As matter we consider Dirac particles of mass~$m$, 
as described by solutions of the Dirac equation~$(\D - m)\Psi =0$. Here the Dirac operator
clearly depends on the metric. On the other hand, the spinors enter the Einstein equations via the
energy momentum tensor of the wave functions. This mutual coupling gives rise to
a system of nonlinear partial differential equations.

In order to keep the equations as simple as possible, we want the whole system to be
again homogeneous and isotropic. This forces us to consider several Dirac particles, as we now
explain in the closed case (the argument in the other cases is similar).
Then the hypersurfaces~$t=\text{const}$ in~\eqref{lineel} are isometric to three-dimensional spheres.
This makes it possible to separate the spatial dependence of~$\Psi$ by the ansatz
 \begin{equation} \label{ansatzs}
\Psi = R(t)^{-\frac{3}{2}}
\left( \begin{matrix} \alpha(t) \:\psi_\lambda(r,\vartheta,\varphi) \\
\beta(t) \:\psi_\lambda(r,\vartheta,\varphi) \end{matrix} \right)
\end{equation}
with complex functions~$\alpha$ and~$\beta$, where~$\psi_\lambda$ is an eigenspinor of the
Dirac operator~$\D_{S^3}$ on~$S^3$
corresponding to the eigenvalue~$\lambda$. The Dirac operator~$\D_{S^3}$ has the
eigenvalues~$\lambda = \pm \frac{3}{2}, \pm \frac{5}{2},\ldots$ with
multiplicities~$N = \lambda^2 - 1/4$. As the symmetry group~$SO(4)$ of~$S^3$ acts
transitively on the $\lambda$-eigenspace, we can retain the $SO(4)$-symmetry
only by taking an anti-symmetric product of all the wave functions which form a basis
of this eigenspace. Thus, similar to the fermion configuration in an inert gas, we fill the
whole ``shell'' of states of momentum~$\lambda$.
The number of particles of our universe equals the dimension~$N = \lambda^2 - 1/4$
of the eigenspace.

We point out that with the ansatz~\eqref{ansatzs}, all one-particle wave functions have the
same time dependence, thus forming a coherent macroscopic quantum state.
Due to the analogy to a BCS state in superconductivity or to the ground state of
a Bose-Einstein condensate, we can thus imagine the fermionic many-particle state as a
{\em{spin condensate}}. The coherence of all wave functions seems essential
for the prevention of space-time singularities. For example, one could consider
an alternative ansatz where one fills several shells corresponding to different
momenta~$\lambda_1, \ldots, \lambda_L$. Apparently, this ansatz would make the
prevention of a singularity less likely. 
Thus the specific features of our model should be attributed to
a spin condensation effect, leading to a coherent macroscopic fermionic state.

With the above ansatz, the Einstein-Dirac equations reduce to ODEs involving the scale
function~$R(t)$ and the above complex functions~$\alpha(t)$ and~$\beta(t)$. Combining~$\alpha$
and~$\beta$ to a two-spinor and taking the expectation values of the Pauli matrices, we can
describe the dynamics of the Dirac wave functions by a vector~$\vec{v} \in \R^3$. After
a suitable orthogonal transformation~$\vec w = U \vec{v}$, the Einstein-Dirac equations take
the simple form
\begin{equation} \label{bloch}
\boxed{ \quad \dot{\vec{w}} = \vec{d} \wedge \vec{w} \:,\qquad
\dot{R}^2+k = -\frac{1}{R^2} \: \sqrt{\lambda^2 + m^2 R^2}\: w_1\:, \quad }
\end{equation}
where the vector~$\vec{d}$ is given by
\begin{equation} \label{ddef}
\vec{d} := \frac{2}{R}\: \sqrt{\lambda^2 + m^2 R^2}\: e_1 \:-\:
 \frac{\lambda m R}{\lambda^2 + m^2 R^2}\: \frac{\dot{R}}{R}\: e_2 \:.
\end{equation}
The~$\vec{w}$-equation in~\eqref{bloch} describes a rotation of the Bloch vector~$\vec{w}$
around the axis~$\vec{d}$, which is itself moving in time. In particular, one sees that the
length of the Bloch vector is constant. We choose the normalization convention
\beq \label{wnorm}
| {\vec w}| = \lambda^2 - \frac{1}{4} = N\:.
\eeq

In the limiting cases~$\lambda=0$ and~$m=0$, our equations reduce to the the well-known Friedmann equations for the {\em dust} universe and the {\em{radiation}} universe, respectively.
Moreover, for large~$R$ the universe behaves classically as in the dust case.
However, near the big crunch or big bang singularities, quantum effects change the situation dramatically.
More precisely, the Dirac equation on the left of~\eqref{bloch} can cause the vector~$\vec{w}$ to
``tilt'', with the effect that~$|w_1|$ in the right equation of \eqref{bloch} can become small.
If this happens, $\dot R$ can become zero even for small values of $R$. As we shall see, in this
case~$\dot{R}$ changes sign, thus preventing the formation of a big bang or big crunch singularity.
We refer to this effect as a {\em{bouncing}} of the scale function.
Our main result is to give a rigorous proof of bouncing and to prove the existence of time-periodic
solutions:
\begin{Thm}[{\bf{Existence of bouncing solutions}}] \label{Thmlimit}
Given~$\lambda \in \{\pm \frac{3}{2}, \pm \frac{5}{2}, \ldots \}$
and $\delta >0$ as well as any radius~$R_>$ and time~$T_>$, there is a
continuous three-parameter family of solutions~$(R(t), \vec{w}(t))$
of the system~\eqref{bloch} defined on a time interval~$[0, T]$ with~$T > T_>$ having the
following properties:
\begin{itemize}
\item[(a)] At~$t=0$ and~$t=T$, the scale function has a local maximum larger than~$R_>$,
\[ R(t) > R_> \:,\qquad \dot{R}(t) = 0 \:,\qquad \ddot{R}(t) < 0\:. \]
\item[(b)] There is a time~$t_\text{bounce} \in (0, T)$ such that~$R$ is strictly monotone
on the intervals~$[0, t_\text{bounce}]$ and~$[t_\text{bounce}, T]$. Moreover, the scale
function becomes small in the sense that
\[ R(t_\text{bounce}) < \delta\, R_>\:. \]
\end{itemize}
\end{Thm} \noindent
By increasing the parameters~$T_>$ and~$R_>$, we can arrange that the
bouncing solution exists for arbitrarily long times and that the scale function becomes
arbitrarily large. With the parameter~$\delta$,
on the other hand, we can make the scale function at the bouncing point as small as we like.
Thus the solutions behave qualitatively as shown in Figure~\ref{fig2} on page~\pageref{fig2}.

\begin{Thm}[{\bf{Existence of time-periodic solutions}}] \label{Thmperiodic}
Given~$\lambda \in \{\pm \frac{3}{2}, \pm \frac{5}{2}, \ldots \}$
and $\delta >0$ as well as any radius~$R_>$ and time~$T_>$, there is a
one-parameter family of solutions~$(R(t), \vec{w}(t))$ of the system~\eqref{bloch}
defined for all~$t \in \R$ with the following properties:
\begin{itemize}
\item[(A)] The solution is periodic, i.e.
\beq \label{percond}
R(t+T) = R(t)\:, \qquad \vec{w}(t+T) = \vec{w}(t) \qquad \text{for all~$t \in \R$}\:,
\eeq
and every~$T>0$ with this property is larger than~$T_>$. \\[-0.5em]
\item[(B)] \hspace*{3.7cm} $\displaystyle{\inf_\R R(t) < \delta \,R_>\:,
\qquad R_> < \sup_\R R(t)}\:.$
\end{itemize}
\end{Thm} \noindent
At first sight, this theorem might seem to be in conflict with the Hawking-Penrose singularity
theorems, which state that a contracting universe must form a space-time singularity.
However, there is no contradiction because our time-periodic solutions violate
the strong energy condition at the minimum of the
scale function (for details see Section~\ref{secenergy}).

The paper is organized as follows. After introducing our model (Section~\ref{sec2}),
in Section~\ref{sec3} we give a qualitative analysis including scalings, a
numerical example, explicit approximate solutions and a discussion of the probability
for bouncing to occur. We proceed with the existence proofs for bouncing solutions
(Section~\ref{existbounce}) and for time-periodic solutions (Section~\ref{existperiodic}).
In Section~\ref{sec6}, we briefly discuss the cases of a flat and open universe. The appendices
provide preliminary and more technical material related to Dirac spinors in curved space-time and
the derivation of the homogeneous and isotropic Einstein-Dirac equations.

\section{The Model} \label{sec2}
The Einstein-Dirac (ED) equations read
\begin{equation}\label{Einst-Dir}
R^i_{j} -\frac 12 R\:\delta^i_j = 8\pi \kappa\,T^i_{j}\:, \qquad (\D
- m) \Psi = 0\:,
\end{equation}
where $T^i_j$ is the energy-momentum tensor of the Dirac particles, $\kappa$ is the
gravitational constant, $\D$ denotes the Dirac operator,
and $\Psi$ is the Dirac wave function (we always work in natural units $\hbar=c=1$).
For the metric we take the ansatz of the spatially homogeneous and isotropic line element
\[ ds^2 = dt^2 - R^2(t)
\left( \frac{dr^2}{f^2(r)} + r^2\,d\vartheta^2 + r^2 \sin^2{\vartheta} \:d\varphi^2 \right) , \]
where~$t \in \R$ is the time of an observer at infinity, $(\vartheta, \varphi) \in (0, \pi) \times [0, 2 \pi)$
are the angular coordinates, and~$r$ is a radial coordinate. The function~$R(t)$ is referred
to as the {\em{scale function}}. In the cases of a closed, open and
flat universe, the function~$f$ and the range of the radial variable~$r$ are given by
\[ \left\{ \begin{array}{cll}
\text{closed universe:} & f(r) = \sqrt{1-r^2} \:,\quad & r \in (0, 1) \\[0.2em]
\text{open universe:} & f(r) = \sqrt{1+r^2} \:,\quad & r >0 \\[0.2em]
\text{flat universe:}  & f(r) = 1 \:,\quad & r >0\:.
\end{array} \right. \]
We combine these formulas by writing
\[ f(r) = \sqrt{1 - k r^2} \qquad \text{with} \qquad k \in \{1, -1, 0\}\:. \]

The Dirac equation in a homogeneous and isotropic space-time is derived in Appendix~\ref{appA}.
By separating the spatial dependence in the form~\eqref{ansatzs},
we obtain a coupled system of ODEs for the complex-valued functions~$\alpha$ and~$\beta$,
\beq \label{DiracODE}
i\, \frac{d}{dt} \left( \begin{matrix} \alpha  \\  \beta \end{matrix} \right)
=  \left( \begin{matrix} m & -\lambda/R \\ -\lambda/R & - m \end{matrix} \right)
\left(\begin{matrix} \alpha \\  \beta \end{matrix}\right) .
\eeq
Here the quantum number~$\lambda$ describes the momentum of the Dirac particle,
whereas~$\psi_\lambda$ is an eigenspinor of the spatial Dirac operator.
In the closed universe, $\lambda$ can take the quantized values~$\pm \frac{3}{2}, \pm \frac{5}{2}, \ldots$
(see~\cite[Appendix~A]{moritz}),
whereas in the flat and open universes, $\lambda$ can be any real number.
We always normalize the spinors according to
\beq \label{normalize}
|\alpha|^2 + |\beta|^2 = \lambda^2 - \frac{1}{4} \:.
\eeq

Note that~$\Psi$ is not homogeneous and isotropic, because its momentum and spin orientation
distinguish a spatial direction, and because its probability density~$|\Psi|^2$ is in general not
constant on the hypersurfaces~$t=\text{const}$. In order to obtain a homogeneous and isotropic
system, we consider a family of wave functions (all of the form~\eqref{ansatzs} with the
same~$\lambda$ and the same~$(\alpha(t), \beta(t))$) and build up a complete fermion shell
(for details see Appendix~\ref{appB}).
A direct calculation yields that the non-vanishing components of the energy-momentum tensor are
(see Appendix~\ref{appB})
\beq \label{Tform} \begin{split}
T^0_0 &= R^{-3} \left[ m \Big(|\alpha|^2-|\beta|^2 \Big)
- \frac{2 \lambda}{R}\: \text{Re}(\alpha \overline{\beta}) \right] \\
T^r_r &= T^\vartheta_\vartheta = T^\varphi_\varphi
= R^{-3} \:\frac{2 \lambda}{3 R}\: \text{Re}(\alpha \overline{\beta}) \:.
\end{split} \eeq
Similarly, a short calculation yields for the Einstein tensor~$G^j_k$
(see~\cite[Box 14.5, eqns~(5)]{misner})
\[ G^0_0 = 3\: \frac{\dot{R}^2+k}{R^2} \:,\qquad
G^r_r = G^\vartheta_\vartheta = G^\varphi_\varphi
=  2 \:\frac{\ddot{R}}{R} + \frac{\dot{R}^2+k}{R^2} \:, \]
and all other components vanish.
Thus the Einstein equations~$G^j_k = 8 \pi \kappa T^j_k$ reduce to the two
ordinary differential equations
\begin{align}
3\: \frac{\dot{R}^2+k}{R^2} &= 8 \pi \kappa\:R^{-3} \left[ m \big(|\alpha|^2-|\beta|^2 \big)
- \frac{2 \lambda}{R}\: \text{Re}(\alpha \overline{\beta}) \right] \label{ein1} \\
6 \:\frac{\ddot{R}}{R} + 6 \:\frac{\dot{R}^2+k}{R^2}
&= 8 \pi \kappa \:R^{-3}\: m \,\big(|\alpha|^2-|\beta|^2 \big) . \label{ein2}
\end{align}
Differentiating the first Einstein equation~\eqref{ein1} and using the Dirac equation~\eqref{DiracODE},
we find that the second Einstein equation~\eqref{ein2} is automatically satisfied.
Thus this equation can be left out. Furthermore, the scaling
\[ R \rightarrow \Lambda R \:,\quad t \rightarrow \Lambda t
\:,\quad m \rightarrow \Lambda^{-1} m \:,\quad \lambda \rightarrow \lambda
\:,\quad (\alpha, \beta) \rightarrow (\alpha, \beta) \:,\quad
\kappa \rightarrow \Lambda^2 \kappa \]
allows us to arbitrarily change the gravitational constant~$\kappa$ (while preserving
the normalization~\eqref{normalize}).
Thus for convenience we may choose
\beq \label{cconvent}
\kappa = \frac{3}{8 \pi} \:.
\eeq
Then the Einstein-Dirac equations become
\beq \label{EinsteinODE}
\dot{R}^2+k = \frac{m}{R} \left(|\alpha|^2-|\beta|^2 \right)
- \frac{2 \lambda}{R^2}\: \text{Re}(\alpha \overline{\beta}) \:.
\eeq

For the analysis of the system of ODEs~\eqref{DiracODE} and~\eqref{EinsteinODE}, it is convenient to
regard the spinor $(\alpha, \beta)$ as a two-level quantum state, and to represent it by a {\em{Bloch
vector}}~$\vec{v}$. More precisely, introducing the $3$-vectors
\beq \label{blochdef}
\vec{v} = \left\langle \!\left( \!\!\begin{array}{c} \alpha \\ \beta \end{array} \!\!\right)\!,
\vec{\sigma} \left( \!\!\begin{array}{c} \alpha \\ \beta \end{array} \!\!\right) \!\right\rangle_{\C^2}
\qquad {\mbox{and}} \qquad
\vec{b} = \frac{2 \lambda}{R}\, e_1 - 2 m e_3
\eeq
(where~$\vec{\sigma}$ are the Pauli matrices, and~$e_1, e_2, e_3$ are the standard basis
vectors in~$\R^3$), the ED equations become
\[ \dot{\vec{v}} = \vec{b} \wedge \vec{v} \:,\qquad
\dot{R}^2+k = -\frac{1}{2R} \: \vec{b} \cdot \vec{v} \]
(where~`$\wedge$' and~`$\cdot$' denote the cross product and the
scalar product in Euclidean~$\R^3$, respectively). To further
simplify the equations, we introduce a rotation~$U$ around the
$e_2$-axis, such that~$\vec{b}$ becomes parallel to~$e_1$,
\beq \label{Udef}
U \vec{b} = \frac{2}{R}\: \sqrt{\lambda^2 + m^2 R^2}\, e_1\:.
\eeq
Then the vector~$\vec{w} := U \vec{v}$ satisfies the equations~\eqref{bloch} and~\eqref{ddef}.
We refer to the two equations in~(\ref{bloch}) as the Dirac and Einstein equations in the
{\em{Bloch representation}}, respectively. Our normalization convention~\eqref{normalize} implies
that the length of the Bloch vector satisfies~\eqref{wnorm}.

\section{Qualitative and Numerical Analysis} \label{sec3}
\subsection{Limiting Cases and Scalings}\label{limcases}
Let us discuss the ED equations in the Bloch representation~\eqref{bloch} and~\eqref{ddef}.
The Dirac equation describes a rotation of the Bloch
vector~$\vec{w}$ around the axis~$\vec{d}$, which is itself moving. 
The direction of the rotation axis is described by the quotient
\beq \label{dquot}
\frac{|d_2|}{|d_1|} = \frac{\lambda  m R\:|\dot{R}|}{2\, (\lambda^2 + m^2 R^2)^\frac{3}{2}} \:.
\eeq
If this quotient is close to zero, the rotation axis is almost parallel to the $e_1$-axis,
so that~$w_1$ is a constant. Conversely, if the quotient is large, then the rotation axis is
parallel to the $e_2$-axis. This has the effect that~$w_1$ has an oscillatory behavior.
This leads to ``quantum oscillations'' of the energy-momentum tensor in the Einstein
equations in~\eqref{bloch}. In physical terms, these oscillations can be understood as the
{\em{Zitterbewegung}} of Dirac particles. In the intermediate regime when the quotient~\eqref{dquot} is
of the order one, the motion of the Bloch vector resembles the precession of a gyroscope. 
We denote the corresponding length scale by~$R_\text{tilt}$,
\beq \label{R1tiltdef}
\frac{|d_2|}{|d_1|} \Big|_{R=R_\text{tilt}} = 1\:.
\eeq

In the regime~$R \gg R_\text{tilt}$ when~$w_1$ is constant, 
the Einstein equation in~\eqref{bloch} decouples from the Dirac equation.
The right side of the Einstein equation in~\eqref{bloch} is a monotone decreasing function of~$R$.
In the case of a closed universe, this implies that~$\dot{R}$ becomes zero at a radius~$R_\text{max}$
determined by
\beq \label{w1det}
-\frac{R_\text{max}^2}{\sqrt{\lambda^2 + m^2 R_\text{max}^2}} = w_1 =: w_{1,{\rm max}} \:.
\eeq
This corresponds to the scale when the expansion area of the universe ends, and a contraction begins.
In the cases of an expanding open or flat universe, it is still useful to introduce~$R_\text{max}$
by~\eqref{w1det}. It is then the length scale where
matter ceases to be the driving force for the dynamics of the universe.

The term~$\lambda^2 + m^2 R^2$ gives rise to yet another length scale, which we denote by~$R_\text{qu}$,
\beq \label{Rqudef}
R_\text{qu} = \frac{\lambda}{m}\:.
\eeq
If~$R \gg R_\text{qu}$ and~$R \gg R_\text{tilt}$, the Einstein equation is well approximated
by the Friedmann equation for {\em{dust}},
\beq \label{dust}
\dot{R}(t)^2+k = \frac{c}{R(t)} \qquad \text{with~$c = - m w_1$} \:.
\eeq
If on the other hand~$R \ll R_\text{qu}$ and~$R \gg R_\text{tilt}$, 
we obtain the Friedmann equation for {\em{radiation}},
\beq \label{radiation}
\dot{R}(t)^2+k = \frac{c}{R(t)^2} 
\qquad \text{with~$c = - \lambda^2 w_1$} \:.
\eeq
The transition from dust to radiation is described by
wave mechanics. This is why~$R_\text{qu}$ can be regarded as the
scale where quantum effects come into play.

Recall that our equations involve the two parameters~$\lambda$ and~$m$.
Moreover, a solution is characterized by the initial orientation of the vector~$\vec{w}$ in~$\R^3$,
giving rise to two additional parameters. Thus our systems admit a four-parameter family of
solutions. Here it is most convenient to parametrize the solution space by the
four quantities
\beq \label{parms}
\Rmax\:, \qquad \lambda \:,\qquad m \qquad \text{and} \qquad
\phi_\text{max} = \arctan \frac{w_3}{w_2} \Big|_{R=R_\text{max}}\:.
\eeq
The parameters~$R_\text{qu}$ and~$\wmax$ are then determined by~\eqref{Rqudef}
and~\eqref{w1det}.
We are interested in the situation when a classical universe gets into the
quantum regime (or similarly if a quantum universe expands into a classical universe).
Therefore, we only consider the parameter regime where
\beq \label{Rineq}
R_\text{max} \gg R_\text{qu}, R_\text{tilt}\:.
\eeq
Then~\eqref{w1det} simplifies to
\beq \label{mrep}
w_{1,{\rm max}} \approx -\frac{R_\text{max}}{m} \:.
\eeq
Let us now follow the solution starting at~$R=R_\text{max}$ through an era of contraction
up to the radius~$R_\text{tilt}$. We can then approximate~$w_1$ by the
constant~$\wmax$. Moreover, for the computation of~$R_\text{tilt}$ we may neglect the
summand~$+k$ in~\eqref{bloch}. Thus we may approximate~\eqref{dquot} by
\begin{align}
\label{c1}
\left| \frac{d_2}{d_1} \right|^2 & = \frac{\lambda^2  m^2 \:|\wmax|}
{2\,(\lambda^2 + m^2 R^2)^\frac{5}{2}} 
\approx  \frac{m \Rmax}{2 \lambda^3} \:
\bigg( 1 + \frac{R^2}{R_\text{qu}^2} \bigg)^{-\frac{5}{2}}\:,
\end{align}
where in the last step we used~\eqref{Rqudef}.
This expression becomes larger if~$R$ decreases, meaning that quantum oscillations become
stronger. Since we want~$R_\text{tilt}$ to exist, we are led to demanding that
\beq \label{Rasy}
\frac{m \Rmax}{\lambda^3} \gtrsim 1\:.
\eeq
Comparing with~\eqref{c1} and~\eqref{R1tiltdef}, we conclude that
\beq\label{Rtilt}
R_\text{qu} \lesssim 
R_\text{tilt} \simeq 
\frac{\lambda^\frac{2}{5} \:\Rqu^{\frac{1}{5}}}{m^\frac{4}{5}} \:.
\eeq
To summarize, we always consider the quantities in~\eqref{parms} as our free parameters,
which we always choose in agreement with~\eqref{Rineq} and~\eqref{Rasy}.
Then the scales~$R_\text{max}$, $R_\text{tilt}$ and~$R_\text{qu}$ automatically comply
with~\eqref{Rtilt}.

\subsection{Construction of Approximate Solutions} \label{approx}
The above considerations yield a method for constructing explicit approximate solutions.
We now describe this method. In Section~\ref{secnumerics}, we shall illustrate it by
a numerical example, while in Section~\ref{prob} we will use it to get information on the
probability of bouncing.
For simplicity, we restrict attention to the closed case~$k=1$, although the methods apply similarly
in the open and flat cases.
In short, we begin at~$t=0$ with~$R(0)=\Rmax$, adjusting parameters such that~$\dot{R}(0)=0$
and~$\ddot{R}(0)<0$. We then approximate the Dirac equation by keeping the rotation axis~$\vec{w}$
fixed in the direction of~$e_1$ (Era~I).
Then~$w_1$ is constant, so that~$R(t)$ goes over to
a solution of the Friedmann equation for dust~\eqref{dust}. We use this approximation up to a
time~$t = \ttilt$ when the scaling function attains the value~$R(\ttilt) = \Rtilt$, where quantum
oscillations become relevant. At the time~$t=\ttilt$, we instantaneously tilt~$\vec{w}$ into the
direction~$e_2$ by omitting the term proportional to~$e_1$ in~\eqref{ddef}.
From then on, we again keep the rotation axis~$\vec{w}$ in the fixed direction~$e_2$
(Era~II). This {\em{approximation of instantaneous tilt}} is exact in the limiting cases~$R \gg \Rtilt$
and~$R \ll \Rtilt$. For the validity in the intermediate region~$R \approx \Rtilt$
see the last paragraph in Section~\ref{secnumerics}.

We again take the quantities in~\eqref{parms} as our free parameters, 
whereas~$R_\text{qu}$ and~$\wmax$ are determined by~\eqref{Rqudef}
and~\eqref{w1det}. In Era~I, we begin at the maximum point $R=\Rmax \gg \Rtilt, \Rqu$.
Thus in the Einstein equation in~\eqref{bloch} we replace~$w_1$ by the constant~$\wmax$.
Using the approximation~\eqref{mrep}, we obtain the initial value problem
\begin{align} \label{phase1eq}
\dot{R}^2 +1 &=  -\frac{m }{R}\: \wmax \:, &
\dot{\vec{w}} &=  2m \: e_1 \wedge \vec w \\ 
R(0) &= \Rmax \:,& \vec w(0) &= \big( \wmax,
\rho \cos (\phi_{\max}) , \rho \sin (\phi_{\max}) \big) \:, \label{phase1init}
\end{align}
where according to our normalization convention~\eqref{wnorm} we
chose
\beq \label{rhodef}
\rho = \sqrt{ N^2-\wmax^2 } \:.
\eeq
The $R$-equation in~\eqref{phase1eq} is the standard Friedmann equation for dust.
It can be solved by separation of variables
and integration (see for example~\cite[page~138]{hawking+ellis}).
The vector $\vec w(t)$ is given explicitly by
\[ \vec w(t) = (\wmax, \rho \cos(2mt + \phi_{\max}), \rho \sin(2mt+ \phi_{\max}))\:. \]
We follow this solution up to the time~$t=\ttilt$ where~$R(\ttilt) = \Rtilt$.
Note that, according to~\eqref{Rtilt}, $R$ always stays larger than~$R_\text{qu}$,
making it unnecessary to consider the equations~\eqref{radiation} for the radiation dominated universe.

In the following Era~II, we approximate the system by
\begin{align} 
\dot{R}^2 +1 &= -\frac{1}{{R}^2} \: \sqrt{\lambda^2 + m^2 {R}^2}\: w_1 , &
\dot{\vec{w}} &=  -\frac{\lambda m \dot R}{\lambda^2 + m^2 R^2}\:  e_2 \wedge \vec w
\label{phase2eq} \\
R(\ttilt) &=\Rtilt \:, & w(\ttilt) &=  \big( \wmax,
\rho \cos (\phi_\text{tilt}) , \rho \sin (\phi_\text{tilt}) \big) , \label{phase2init}
\end{align}
where we set
\beq \label{phitilt}
\phi_\text{tilt} = 2m \,\ttilt + \phi_{\max}\:.
\eeq
As the $\vec{w}$-equation in~\eqref{phase2eq} describes a rotation around
the~$e_2$-axis, the component~$w_2$ is constant. Therefore, it suffices to consider
the $(e_1, e_3)$-plane. We describe the rotation of the vector~$(w_1, w_3)$ in this plane
by an angle~$\theta$, i.e.
\beq \label{thetapar}
\begin{pmatrix} w_1(t) \\ w_3(t) \end{pmatrix} = 
\begin{pmatrix} \cos \theta & \sin \theta \\
-\sin \theta & \cos \theta \end{pmatrix} \begin{pmatrix} w_1(\ttilt) \\ w_3(\ttilt) \end{pmatrix} \:.
\eeq
Then the $\vec{w}$-equation can be rewritten as 
\begin{equation}\label{theta}
\dot \theta = - \frac d{dt} \arctan \bigg[ \frac{R}{\Rqu} \bigg] , \qquad
\theta(\ttilt) = 0,
\end{equation}
having the explicit solution
\beq \label{thetaex}
\theta(R) = \arctan \bigg[ \frac{\Rtilt}{\Rqu} \bigg] - \arctan \bigg[ \frac{R}{\Rqu} \bigg]\:.
\eeq
This can now be used in~\eqref{phase2eq} to obtain the differential equation
\beq \label{phase2Rtheta}
\dot{R}^2 + 1 = -\frac{\sqrt{\lambda^2 + m^2 {R}^2}}{{R}^2} \: \Big( \wmax \cos \theta(t) 
+\rho \sin(\phi_\text{tilt}) \sin \theta(t) \Big) .
\eeq
Clearly, this nonlinear equation has a local solution with initial
condition~$R(\ttilt)=\Rtilt$, which can be maximally extended until
$1/R(t)$ diverges at the moment of a big crunch. 

\subsection{Numerical Results} \label{secnumerics}
\begin{figure}
\includegraphics[width=6.5cm]{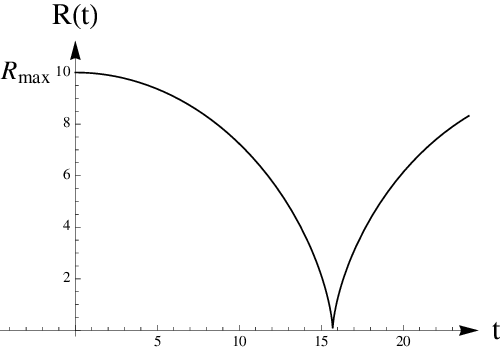} $\qquad$
\includegraphics[width=6.5cm]{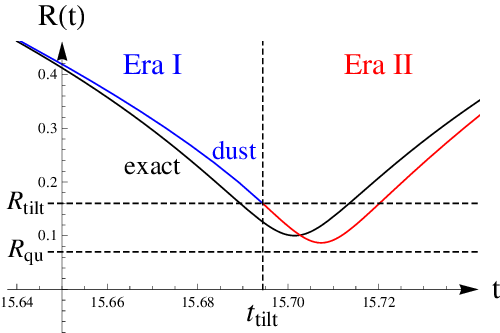} \\[1.5em]
\includegraphics[width=6.5cm]{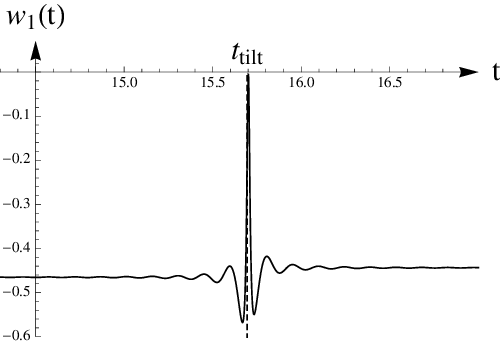} $\qquad$
\includegraphics[width=6.5cm]{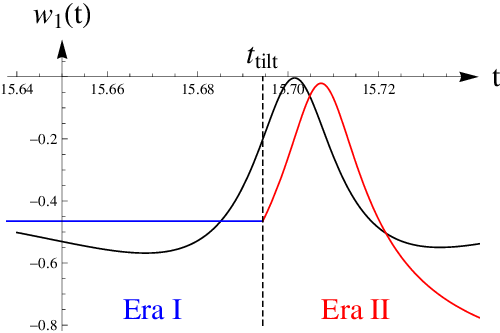}
\caption{The exact solution and the approximation of immediate tilt
in the case~$m=21.5$, $\lambda=3/2$, $R_{\max}=10$, $R_\text{tilt} \approx 0.160$, $R_\text{qu}
\approx 0.0697$, $\wmax \approx -0.465$.}
 \label{figbounce}
\end{figure}

We now consider the approximation of instantaneous tilt in a typical example. On
the left of Figure~\ref{figbounce},
a numerical solution of the ED equations~\eqref{bloch} and~\eqref{ddef} is plotted, where the
universe begins in an era of classical contraction. At~$t \approx 15$, quantum oscillations come into
play, which trigger a bouncing of the scale function leading to a new era of classical expansion.
The function~$w_1$ is almost constant in the classical eras, but is oscillatory in the quantum regime.
In the approximation of instantaneous tilt, $w_1$ is exactly constant up to the time~$t_\text{tilt}$
and oscillates afterwards. As one sees in Figure~\ref{figbounce}, this is a quite
good approximation (in the upper left plot for~$R(t)$, the approximate solution
looks just like the exact solution until after the bouncing).

Clearly, the limitation of the approximation of instantaneous tilt is that it does not take into account
the ``precession'' of the Bloch vector around the moving axis of rotation~$\vec{w}$ in the
region~$R \approx R_\text{tilt}$. 
Generally speaking, the approximation is better if the expression on the left side of~\eqref{Rasy} is larger
(in the example of Figure~\ref{figbounce}, this expression has the value $\approx 63.7$).
A precise error estimate goes beyond the scope of this paper, because it sees difficult
to describe the above precession quantitatively.

\subsection{The Probability of Bouncing}\label{prob}
As the phase~$\phi_\text{tilt}$ changes fast during the era of classical contraction (see~\eqref{phitilt}),
it is reasonable to assume that this phase is random and uniformly distributed 
in $[0,2\pi) \!\!\mod 2 \pi$. Under this assumption, we can compute the probability of bouncing
within the approximation of instantaneous tilt. Namely, a bouncing occurs if~$\dot{R}$ vanishes
for some time~$t > \ttilt$. According to~\eqref{phase2Rtheta}, this is equivalent to the condition
that there is~$R \in [0, \Rtilt)$ such that
\beq \label{bcond0}
\frac{R^2}{\sqrt{\lambda^2 + m^2 R^2}} = - \wmax \cos \theta(R)
- \rho \sin(\phi_\text{tilt}) \sin \theta(R)
\eeq
with~$\theta(R)$ as given by~\eqref{thetaex}. This condition can easily by evaluated numerically.
Moreover, we obtain the sufficient condition for bouncing that the right side of~\eqref{bcond0} is
negative at~$R=0$, i.e.
\beq \label{bcond}
\sin \phi_\text{tilt} > \frac{R_\text{qu}\: |\wmax|}{\Rtilt \:\rho} \:,
\eeq
where we used that according to~\eqref{thetaex}, $\tan \theta(0) = R_\text{tilt}/R_\text{qu}$.
If the right side of~\eqref{bcond} is smaller than one, we can estimate the probability~$p$ of bouncing
by
\[ p \geq \frac{1}{\pi} \: \arccos \left( \frac{R_\text{qu}\: |\wmax|}{\Rtilt \:\rho} \right) \:. \]
Here~$\wmax$ and~$\rho$ can be rewritten using~\eqref{w1det} or~\eqref{mrep}
and~\eqref{rhodef}.

\section{Existence of Bouncing Solutions} \label{existbounce}
In the previous sections we presented numerical solutions together with approximate analytical arguments
which indicated that quantum effects can lead to a bouncing of the scale function, thus preventing
the formation of a space-time singularity. We shall now prove that such bouncing solutions exist.
The basic strategy is to begin at~$t=0$ at a minimum of the scale function,
\beq \label{Rmin}
R(0) = R_I \:,\qquad \dot R(0) = 0 \:,\qquad \ddot R(0) > 0 \:,
\eeq
which we choose to be microscopic in the sense that~$R_I \ll \Rtilt$.
This solution describes an expanding universe. We show that the solution exists all the way
to the classical era where it goes over to a Friedmann dust solution.

In order to describe the solution and the initial conditions near the minimum~\eqref{Rmin},
it is useful to rescale the solution and the parameters as follows,
\beq \label{scal1}
m \to \frac{m}{\eps},  \quad t\to \frac{t}{\eps^2}\:,\quad
R(t) \to \eps R(t/\varepsilon^2)\:, \quad \lambda \to \lambda, \quad \vec w(t) \to \vec w(t/\eps^2)
\eeq
(note that this scaling leaves~$\lambda$ as well as the length of the Bloch vector
unchanged, so that the normalization condition~\eqref{wnorm} is preserved).
The corresponding {\em{rescaled Bloch equations}} are given by
\beq \label{rescbloch}
\begin{split}
\dot{R}_\eps^2 +\eps^2&= -\frac{1}{R^2} \: \sqrt{\lambda^2 + m^2 R_\eps^2}\: (w_\eps)_1, \\
\dot{\vec{w}}_\eps &=   \left( \eps \:\frac{2}{R_\eps}\:
\sqrt{\lambda^2 + m^2 R_\eps^2}\: e_1 \:-\:
\frac{\lambda m \dot{R}_\eps}{\lambda^2 + m^2 R_\eps^2}\:  e_2  \right) \wedge \vec w_\eps \:,
\end{split}
\eeq
where for clarity, we denoted the solutions by a subscript~$\varepsilon$. 
In the limit~$\varepsilon \searrow 0$, these equations go over to the
so-called {\em{microscopic limit equations}}
\beq \label{blochlimit}
\dot{R}^2 = -\frac{1}{{R}^2} \: \sqrt{\lambda^2 + m^2 {R}^2}\: w_1\:, \qquad
\dot{\vec{w}} = -\frac{\lambda m \dot R}{\lambda^2 + m^2 R^2}\:  e_2 \wedge \vec w\:.
\eeq
We remark that that these equations are very similar to the equations~\eqref{phase2eq}
used in the approximation of instantaneous tilt. The only difference is the additional
summand~$+1$ in the~$R$-equation~\eqref{phase2eq}. Also, one should keep in mind
that the microscopic limit equations involve the rescaling~\eqref{scal1}, whereas the
equations~\eqref{phase2eq} merely are an approximation of the original system~\eqref{bloch}.

\subsection{Solution of the Microscopic Limit Equations} \label{secmle}
We now solve the approximate limit equations for the initial values~\eqref{Rmin}.
First, one sees from the $R$-equation in~\eqref{blochlimit} that~$w_1(0)=0$.
Next, the $\vec{w}$-equation in~\eqref{blochlimit} yields that~$w_2$ is constant.
Parametrizing~$w_1$ and~$w_3$ similar to~\eqref{thetapar} by an angle~$\Theta$, we thus obtain
\beq \label{mle1}
w_2(t) \equiv w_2(0) \:,\qquad \begin{pmatrix} w_1(t) \\ w_3(t) \end{pmatrix} = 
\begin{pmatrix} \sin \theta \\
\cos \theta \end{pmatrix} w_3(0) \quad \text{with} \quad \theta|_{t=0}=0\:.
\eeq
Exactly as described after~\eqref{thetapar}, the $\vec{w}$-equation
can now be solved explicitly,
\beq \label{thetamle}
\theta(R) = \arctan \bigg[ \frac{R_I}{\Rqu} \bigg] - \arctan \bigg[ \frac{R}{\Rqu} \bigg]\:.
\eeq
In particular, one sees that~$\theta(R)<0$ for all~$R > R_I$, so that~$w_1(t)$ has the opposite
sign as~$w_3(0)$.
Thus in order to get consistency with the $R$-equation in~\eqref{blochlimit}, we must
choose
\beq \label{w3sign}
w_3(0)>0 \:.
\eeq
Then our initial conditions indeed describe a minimum of the
scale function~\eqref{Rmin}. More specifically, the $R$-equation becomes
\[ \dot{R}^2 = -\frac{\sqrt{\lambda^2 + m^2 {R}^2}}{{R}^2} \:
\sin (\Theta(R)) \: w_3(0)\:, \]
which can be solved by separation,
\[ t = \int_{R_I}^R \frac{dR}{\dot{R}} =
\int_{R_I}^R \bigg( -\frac{\sqrt{\lambda^2 + m^2 {R}^2}}{{R}^2} \:
\sin \big( \Theta(R) \big) \: w_3(0) \bigg)^{-\frac{1}{2}} \,dR \:. \]
This equation is useful for analyzing the behavior as~$R \rightarrow \infty$.
Namely, the expansion
\[ -\frac{\sqrt{\lambda^2 + m^2 {R}^2}}{{R}^2} \:
\sin (\Theta(R)) \: w_3(0) = \pm
\frac{m}{R}\: w_3(0)\: \sin \arctan \bigg[ \frac{R_I}{\Rqu}\bigg]  + {\mathcal{O}} \Big(\frac{1}{R^2} \Big) \]
shows that
\beq \label{tRscale}
t(R) \sim \pm \frac{R^\frac{3}{2}}{\sqrt{m}} + {\mathcal{O}} \big( \sqrt{R} \big) \:.
\eeq
In particular, our solutions of the microscopic limit equations exist for all times.

\subsection{Construction of Bouncing Solutions} \label{secbconstruct}
We now turn attention to the rescaled equations~\eqref{rescbloch}
(for fixed~$\lambda, m > 0$). The continuous dependence of solutions of ODEs
on the coefficients immediately gives us the following result.
\begin{Lemma} \label{lemmabs1} Let~$(R_0, \vec{w}_0)$ be a solution of the
microscopic limit equations~\eqref{blochlimit} with initial conditions~\eqref{Rmin}
and~\eqref{mle1}. Then for every~$T>0$ and~$\delta>0$ there is~$\varepsilon>0$
and a family of solutions~$(R_\varepsilon, \vec{w}_\varepsilon)$ of the
rescaled equations~\eqref{rescbloch} such that for all~$t \in [-T,T]$,
\[ |R_\eps(t) - R_0(t)| \leq \delta\:, \qquad |\vec{w}_\eps(t) - \vec{w}(t)| \leq \delta \:. \]
\end{Lemma}

As we saw in Section~\ref{secmle}, the solution~$R_0(t)$ exists for all times,
is strictly monotone on the time intervals~$(-\infty, 0]$ and~$[0, \infty)$ and
diverges in the limit~$t \rightarrow \infty$. The function~$R_\varepsilon$, however,
will not be strictly monotone for all positive times. This is obvious from the fact that the right side
of the $R$-equation in~\eqref{rescbloch} tends to zero in the limit~$R \rightarrow \infty$.
Thus there must be a minimal time~$T_{\max}(\eps, R_I)$ when~$\dot{R}_\varepsilon$ vanishes.
Then clearly~$\dot{R}_\varepsilon$ does not change signs on the interval~$[0, T_{\max}]$.
Moreover, possibly by further decreasing~$\eps$ and choosing~$\delta$ sufficiently
small, we can arrange in view of Lemma~\ref{lemmabs1}
that~$T_{\max}(\eps, R_I)> T$.
Next, as the angle~$\theta_0(R)$ is negative and decreasing according to~\eqref{thetamle},
we can choose~$t_0>0$, $\kappa \in (0,1)$ and a corresponding~$r_0 := R_0(t_0)$ such that
$(w_0)_1(R) < -3 \kappa$ for all~$R>r_0$. Then for any~$T>t_0$, Lemma~\ref{lemmabs1}
ensures that by choosing~$\delta>0$ sufficiently small, we can arrange that~$(w_\varepsilon)_1(T)
< -2\kappa$. The following lemma shows that by increasing~$T$ and further decreasing~$\delta$,
we can even arrange that
\beq \label{w1bound}
(w_\varepsilon)_1(t) < -\kappa < 0  \quad \text{for all~$t \in [T, T_{\max}]$
and all sufficiently small~$\varepsilon>0$}\:.
\eeq
\begin{Lemma} 
Let~$(R_\eps, \vec{w}_\eps)$ be a solution of the rescaled Bloch equations~\eqref{rescbloch}
defined on a time interval~$[T_0,T_1]$.
Assume that~$\dot R_\eps$ does not change signs on this interval. Then
\begin{align}
\bigg| \arcsin \Big( \frac{(w_\eps)_1(T_1)}{N} \Big) &-\arcsin \Big(
\frac{(w_\eps)_1(T_0)}{N} \Big) \bigg| \nonumber \\
& \leq \big| \arctan[R_\eps(T_1) / \Rqu] - \arctan[R_\eps(T_0) / \Rqu] \big| \label{arcsin}
\end{align}
(where~$N=|\vec{w}|$ again denotes the number of particles).
\end{Lemma}
\begin{proof} For ease in notation, we omit the subscript~$\varepsilon$.
The first component of the $\vec{w}$-equation in~\eqref{rescbloch} reads
\[ \dot{w}_1 = - \frac{\lambda m \dot R}{\lambda^2 + m^2 R^2}\: w_3 \:. \]
Employing the inequality~$w_3 \leq \sqrt{N^2-w_1^2}= N \sqrt{1 - \left(w_1/N \right)^2}$,
we obtain the estimate
\[ \left| \frac{\dot w_1}N \right| \leq \bigg| \frac{\lambda m \dot R}{\lambda^2 + m^2 R^2}\bigg|\:
\sqrt{1-(w_1/N)^2}\:, \]
and thus
\[ \left| \frac{d}{dt}  \arcsin({w}_1(t)/N) \right| \leq \left|\frac{d}{dt} \arctan[R(t)/\Rqu] \right|
= \left|\frac{d}{dt} \Big| \!\arctan[R(t)/\Rqu] \Big| \right|, \]
where in the last step we used that~$\dot{R}$ has a fixed sign.
\end{proof} \noindent
The next lemma shows that, possibly by further increasing~$T$ and decreasing~$\delta$,
we can arrange that this~$T_{\max}$ is indeed a local maximum of~$R_\varepsilon$.
\begin{Lemma} \label{lemmaddotR}
Assume that~\eqref{w1bound} holds and that there is~$t>T$ such that
\beq \label{tcond}
R_\eps(t) > \frac{\Rqu \,N}{\kappa} \qquad \text{and} \qquad \dot{R}_\eps(t)=0 \:.
\eeq
Then~$\ddot{R}_\varepsilon(t) \neq 0$.
\end{Lemma}
\Proof We only consider the original ED equations, because then the result is immediately
obtained by rescaling according to~\eqref{scal1}.
Assume that that statement of the lemma is false. Then there is~$t>T$ such that~\eqref{w1bound}
and~\eqref{tcond} hold but~$\ddot{R}(t)=0$.
Combining~\eqref{ein1} with~\eqref{ein2} and using~\eqref{cconvent}, we find
\beq \label{RddotR}
2 R \ddot{R} = -2\, (\dot{R}^2 + 1) + \frac{m}{R}\: (|\alpha|^2 - |\beta|^2) 
\eeq
Again using~\eqref{ein1}, we conclude that at~$t$,
\[ \frac{m}{R}\: (|\alpha|^2 - |\beta|^2) = 2 \qquad \text{and} \qquad
\frac{2 \lambda}{R^2}\: \text{Re}(\alpha \overline{\beta}) = 1 \:. \]
Using~\eqref{blochdef}, we obtain for the Bloch vector~$\vec{v}$ that
\[ v_3 = \frac{2 R}{m} \qquad \text{and} \qquad
v_1 = \frac{R^2}{\lambda} \:. \]
Computing the corresponding vector~$\vec{w} = U \vec{v}$
with~$U$ as given by~\eqref{Udef}, we find that
\[ w_1 = -\frac{1}{\sqrt{R^2+\Rqu^2}}\: \frac{R^2}{m} \qquad \text{and} \qquad
w_3 = \frac{R}{\sqrt{R^2+\Rqu^2}}\: \frac{R^2 + 2 \Rqu^2}{\lambda} \:. \]
It follows that
\[ \frac{|w_1|}{N} \leq \frac{|w_1|}{|w_3|} = \frac{R^2 \lambda}{mR\, (R^2 + 4 \Rqu^2)}
\leq \frac{\lambda}{mR} = \frac{\Rqu}{R}\:, \]
in contradiction to our assumptions~\eqref{w1bound} and~\eqref{tcond}.
\QED \noindent
Arguing similarly for negative times, we find that the scale function also has a maximum
at some time~$T_{\min} < -T$. We have thus proved the following result.
\begin{Lemma} \label{lemmasummary}
For every sufficiently large~$T$ and every sufficiently small~$\varepsilon>0$,
there is a solution~$(R_\varepsilon, \vec{w}_\varepsilon)$ of the rescaled Bloch
equations~\eqref{rescbloch} defined on the interval~$[T_{\min}, T_{\max}]$
with~$T_{\min}<-T$ and~$T_{\max}>T$ with the following properties:
At times~$t=T_{\max}$ and~$t=T_{\min}$, the scale function has a local maximum,
\[ \dot{R}_\varepsilon(t) = 0\:,\qquad \ddot{R}_\varepsilon(t) < 0\:. \]
Moreover, the scale function~$R_\varepsilon$ is monotone decreasing on~$[T_{\min}, 0]$
and monotone increasing on~$[0,T_{\max}]$.
\end{Lemma}

Now we can complete the proof of Theorem~\ref{Thmlimit}.
After performing the scaling~\eqref{scal1} backwards, we obtain a solution~$(R, \vec{w})$
of the original equations~\eqref{bloch}. By decreasing~$\varepsilon$ or increasing~$T$,
we can make the scale function and the total time of the contraction and expansion cycle
as large as we want. We finally note that for fixed~$\lambda$, our construction involves
the free parameters~$R_I$, $w_3>0$ as well as any sufficiently small~$\varepsilon>0$, giving
rise to a continuous three-parameter family of solutions. This proves Theorem~\ref{Thmlimit}.

We conclude this section by analyzing the solutions~$(R_\varepsilon(t), \vec{w}_\varepsilon(t))$
of Lemma~\ref{lemmasummary} outside the time interval~$[-T,T]$ in more detail. In particular, we
want to show that the solutions go over to the Friedmann dust solution as~$|t|$ gets large.

\begin{Prp} \label{lemmaconcave}
The function~$R_\varepsilon^\frac{3}{2}$ is concave if
\[ R_\varepsilon > \frac{\sqrt{\lambda N}}{\varepsilon} \:. \]
\end{Prp}
\Proof 
Again, we only consider the unscaled ED equations, because then the result
follows by rescaling according to~\eqref{scal1}.
Using~\eqref{ein1}, we rewrite~\eqref{RddotR} as
\beq \label{RddR}
2 R \ddot{R} = -(\dot{R}^2 + 1) + \frac{2 \lambda}{R^2}\: \text{Re}(\alpha \overline{\beta}) \:.
\eeq
It follows that
\[ \frac{d^2}{dt^2} R^\frac{3}{2} = \frac{3}{4}\: R^{-\frac{1}{2}} \left( 2 R \ddot{R} + \dot{R}^2 \right)
= \frac{3}{4}\: R^{-\frac{1}{2}} \Big(
-1 + \frac{2 \lambda}{R^2}\: \text{Re}(\alpha \overline{\beta}) \Big) . \]
According to the definition of the Bloch vectors~\eqref{blochdef} and our normalization
convention~\eqref{wnorm},
\[ |\alpha|^2 + |\beta|^2 = |\vec{v}| = |\vec{w}| = N \:. \]
We conclude that
\[ \frac{d^2}{dt^2} R^\frac{3}{2} \leq \frac{3}{4}\: R^{-\frac{1}{2}} 
\Big( -1 + \frac{\lambda N}{R^2} \Big) , \]
showing that~$R^\frac{3}{2}$ is indeed concave if~$R^2 > \lambda N$.
\QED
Clearly, this statement implies in particular that~$R_\varepsilon$ itself becomes concave.

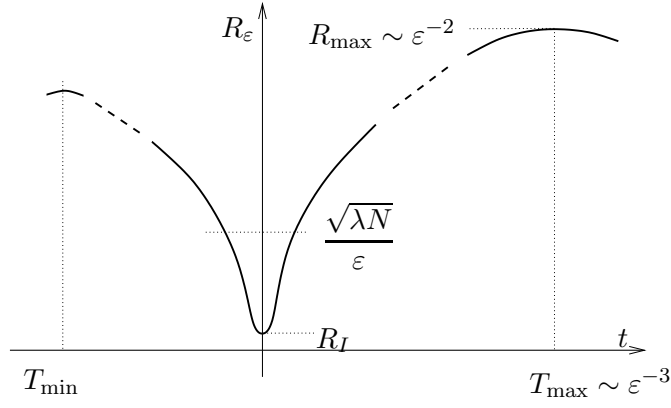
\begin{figure}
\begin{picture}(0,0)%
\includegraphics{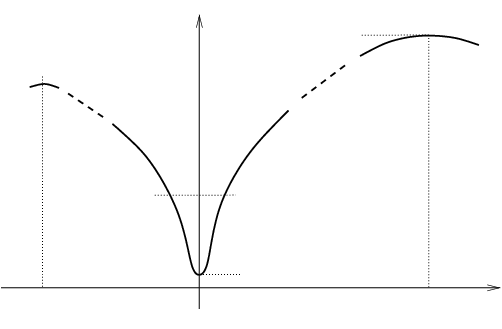}%
\end{picture}%
\setlength{\unitlength}{1575sp}%
\begingroup\makeatletter\ifx\SetFigFontNFSS\undefined%
\gdef\SetFigFontNFSS#1#2#3#4#5{%
  \reset@font\fontsize{#1}{#2pt}%
  \fontfamily{#3}\fontseries{#4}\fontshape{#5}%
  \selectfont}%
\fi\endgroup%
\begin{picture}(10049,6176)(-621,-7925)
\put(-382,-7718){\makebox(0,0)[lb]{\smash{{\SetFigFontNFSS{11}{13.2}{\familydefault}{\mddefault}{\updefault}$T_{\min}$}}}}
\put(4313,-5506){\makebox(0,0)[lb]{\smash{{\SetFigFontNFSS{11}{13.2}{\familydefault}{\mddefault}{\updefault}$\displaystyle{\frac{\sqrt{\lambda N}}{\varepsilon}}$}}}}
\put(4098,-2274){\makebox(0,0)[lb]{\smash{{\SetFigFontNFSS{11}{13.2}{\familydefault}{\mddefault}{\updefault}$R_{\max} \sim \varepsilon^{-2}$ }}}}
\put(7583,-7771){\makebox(0,0)[lb]{\smash{{\SetFigFontNFSS{11}{13.2}{\familydefault}{\mddefault}{\updefault}$T_{\max} \sim \varepsilon^{-3}$}}}}
\put(8975,-7046){\makebox(0,0)[lb]{\smash{{\SetFigFontNFSS{11}{13.2}{\familydefault}{\mddefault}{\updefault}$t$}}}}
\put(2731,-2161){\makebox(0,0)[lb]{\smash{{\SetFigFontNFSS{11}{13.2}{\familydefault}{\mddefault}{\updefault}$R_\varepsilon$}}}}
\put(4220,-7061){\makebox(0,0)[lb]{\smash{{\SetFigFontNFSS{11}{13.2}{\familydefault}{\mddefault}{\updefault}$R_I$}}}}
\end{picture}%
\caption{The bouncing solution~$R_\varepsilon$.}
 \label{fig2}
\end{figure}

Now we can estimate the time~$T_\text{max}(\varepsilon, R_I)$ and the corresponding
value of~$R_\varepsilon$ from above and below.
\begin{Prp} \label{prplower} There are constants $c_1, c_2, c_3, c_4>0$ such that
for all sufficiently small~$\varepsilon>0$,
\beq\label{Tmaxest}
\frac{c_1}{\eps^3} \leq T_{\rm max}(\eps,R_I) \leq \frac{c_2}{\eps^3}
\eeq
\beq\label{Rmaxest}
\frac{c_3}{\eps^2} \leq \Rmax := R \big( T_{\max}(\eps,R_I) \big) \leq \frac{c_4}{\eps^2}\:.
\eeq
\end{Prp}
\begin{proof}
We choose~$T$ so large that the assumptions of Lemma~\ref{lemmaddotR} are satisfied.
Then according to~\eqref{w1bound} and~\eqref{wnorm},
\beq \label{w1bound2}
-N \leq (w_\varepsilon)_1(t) \leq -\kappa \qquad \text{for all~$t>T$}
\eeq
and all sufficiently small~$\varepsilon$.
Using these inequalities in the $R$-equation in~\eqref{phase2eq} and setting~$\dot{R}(T_{\max})=0$,
we obtain~\eqref{Rmaxest}.

In order to derive~\eqref{Tmaxest}, we note that obviously $R_\eps(t) \leq \tilde R(t)$, where~$\tilde{R}$
is the solution of the initial value problem
\[ {\dot {\tilde R}}^2 = \frac{N}{{\tilde R}^2} \: \sqrt{\lambda^2 + m^2 {\tilde R}^2}, \qquad
\tilde R(T) = R_\eps(T) \:. \]
By explicit integration we find that~$\tilde R \leq c t^{2/3}$ for some $c>0$.
It follows that
\[ c { T_{\rm max}(\eps,R_I)}^{2/3} \geq \tilde R(T_{\rm max}(\eps,R_I)) \geq R_\eps( T_{\rm max}(\eps,R_I)) \geq c_3 /\eps^2 \:, \]
giving the lower bound in~\eqref{Tmaxest}. The upper bound follows similarly
from the inequality~$R_\eps(t) \leq \tilde R(t)$ where now~$\tilde{R}$ is the solution
the initial value problem
\[ {\dot{\tilde R}}^2 = \frac{\kappa}{{\tilde R}^2} \: \sqrt{\lambda^2 + m^2 {\tilde R}^2}
, \qquad \tilde R(T) = R_\eps(T) \:. \]

\vspace*{-1.9em}
\end{proof} \noindent
Note that the scalings in Proposition~\ref{prplower} coincide with those of the Friedmann
dust solution. This quantifies that for large times, our ED universe indeed behaves classically.
Our constructions are illustrated in Figure~\ref{fig2}.

\subsection{The Energy Conditions} \label{secenergy}
Let us analyze whether our bouncing solutions satisfy the energy conditions
as described in~\cite[Section~4.3]{hawking+ellis}. 
We first bring the energy conditions into a form suitable for our analysis.
Recall that the {\em{weak energy condition}} demands that~$T_{ab} W^a W^b \geq 0$
for any timelike vector~$W \in T_p M$. Likewise, the {\em{dominant energy condition}} states that
$T_{ab} W^a W^b \geq 0$ and that~$T^{ab} W_a$ is a non-spacelike vector.
Finally, the {\em{strong energy condition}} states that~$R_{ab} W^a W^b \geq 0$.
Due to homogeneity, it here suffices to evaluate these conditions at a space-time point~$p$ with
radius~$r=0$. Moreover, due to isotropy, we can assume that the spatial part of~$W$ points into
the radial direction, i.e.\
\[ W = \tau \partial_t + \rho \partial_r \qquad
\text{with} \qquad 0 \leq g_{ab} W^a W^b = \tau^2 - R^2 \: \rho^2\:. \]
Then
\begin{align*}
T_{ab} W^a W^b &= T^0_0 \tau^2 - T^r_r \: R^2 \rho^2 \\
g_{ac} \:(T^a_b \,W^b) \:(T^c_d \,W^d) &=
\tau^2 (T^0_0)^2 -  R^2 \rho^2 \:(T^r_r)^2\:,
\end{align*}
whereas the components of the Ricci tensor are computed by
\begin{align*}
T &= T^a_a = T^0_0 + 3 \,T^r_r \\
R^a_b &= 8 \pi \kappa \left( T^a_b - \frac{1}{2}\: T\: \delta^a_b \right) \\
R^0_0 &= 4 \pi \kappa \left( T^0_0 - 3 \,T^r_r \right) \\
R^r_r &= 4 \pi \kappa \left( -T^0_0 - T^r_r \right) .
\end{align*}
Using these formulas, the energy conditions become
\beq \label{energycond}
\left\{ \begin{array}{ccl} 
\text{weak energy condition:} &\quad& T^0_0 \geq \max \big(0, T^r_r \big) \\[0.3em]
\text{dominant energy condition:} && T^0_0 \geq |T^r_r| \\[0.3em]
\text{strong energy condition:} && T^0_0 \geq \max \big( 3 \,T^r_r, T^r_r \big) .
\end{array} \right.
\eeq

Using~\eqref{blochdef}, we can express~\eqref{Tform}
in terms of the Bloch vector~$\vec{v}$,
\[ T^0_0 = R^{-3} \left( m \:v_3
- \frac{\lambda}{R}\: v_1 \right) \:,\qquad
T^r_r = R^{-3} \:\frac{\lambda}{3 R}\: v_1 \:. \]
Computing the vector~$\vec{w} = U \vec{v}$ as in the proof of Lemma~\ref{lemmaddotR},
we obtain
\beq \label{TB1}
\begin{split}
T^0_0 &= -\frac{m}{R^4}\: \sqrt{R^2+\Rqu^2} \:w_1  \\
T^r_r &= \frac{\lambda^2}{3 m\, R^4}\: \frac{1}{\sqrt{R^2+\Rqu^2}}\: w_1
+ \frac{\lambda}{3 R^3}\:\frac{1}{\sqrt{R^2+\Rqu^2}}\: w_3 \:.
\end{split}
\eeq
Rescaling by~\eqref{scal1} only gives an overall factor of~$\varepsilon^{-4}$,
\beq \label{TB2}
\begin{split}
\varepsilon^4 \:T^0_0 &= -\frac{m}{R_\varepsilon^4}\: \sqrt{R_\varepsilon^2+\Rqu^2} \:(w_\eps)_1  \\
\varepsilon^4 \:T^r_r &= \frac{\lambda^2}{3 m\, R_\varepsilon^4}\: \frac{1}{\sqrt{R_\varepsilon^2
+\Rqu^2}}\: (w_\eps)_1 + \frac{\lambda}{3 R_\varepsilon^3}\:\frac{1}{\sqrt{R_\varepsilon^2+\Rqu^2}}\: (w_\eps)_3 \:.
\end{split}
\eeq

We first show that the energy conditions are satisfied for large~$R_\varepsilon$, as
was to be expected in view of the Friedmann dust limit.
\begin{Prp} Assume that~\eqref{w1bound} holds and that
\beq \label{Repsbound}
R_\varepsilon \geq \frac{2 N \Rqu}{\kappa} \:.
\eeq
Then the weak, dominant and strong energy conditions are satisfied.
\end{Prp}
\Proof We again only consider the unscaled equations.
Using~\eqref{wnorm} and~\eqref{w1bound}, we can estimate~\eqref{TB1} by
\[ T^0_0 \geq -\frac{m}{R^3}\: w_1 \geq \frac{m \kappa}{R^3} \:,\qquad
3\,|T^r_r| \geq \frac{m}{R^3} \frac{\Rqu^2}{R^2} \: N + \frac{\lambda}{R^4}\: N \:. \]
Hence
\[ T^0_0 - 3\,|T^r_r| \geq \frac{m}{R^3} \Big( \kappa - \frac{\Rqu^2}{R^2} \: N
- \frac{\Rqu}{R}\: N \Big) \geq 0\:, \]
where in the last step we applied~\eqref{Repsbound}.
\QED

A more interesting question is whether the energy conditions are satisfied in the
bouncing regime. We first prove our results and discuss them afterwards.
\begin{Prp} \label{prpsec}
Suppose that a solution bounces at time~$t$, i.e.\
\[ \dot{R}_\varepsilon(t) =0 \qquad \text{and} \qquad
\ddot{R}_\varepsilon(t) > 0\:. \]
Then the strong energy condition is violated at~$t$.
\end{Prp}
\Proof We again consider the unscaled equations.
Comparing~\eqref{TB1} with~\eqref{bloch}, one sees that
\[ T^0_0 = \frac{1}{R^2}\: (\dot{R}^2+1) = \frac{1}{R^2} \:. \]
Moreover, combining~\eqref{RddR} with~\eqref{Tform}, we find
\[ 0 < 2 R \dot{R} = -1 + \frac{2 \lambda}{R^2}\: \text{Re}(\alpha \overline{\beta}) 
= -1 + 3R^2\: T^r_r \]
and thus
\[ T^0_0 - 3 T^r_r < \frac{1}{R^2} - 3\, \frac{1}{3R^2} = 0 \:, \]
in contradiction to the strong energy condition in~\eqref{energycond}.
\QED

\begin{Prp} \label{prpeclimit}
Let~$(R_\varepsilon, \vec{w}_\varepsilon)$ be a the solutions of Lemma~\ref{lemmasummary}.
Then for sufficiently small~$\varepsilon>0$,
the weak, the dominant and the strong energy conditions are violated at the bouncing time.
\end{Prp}
\Proof Suppose that~$\dot{R}_\varepsilon(t)=0$. Solving the $R_\varepsilon$-equation
in~\eqref{rescbloch} for~$(w_\varepsilon)_1$ and substituting the result into~\eqref{TB2},
we find that
\begin{align*}
T^0_0 &= -\varepsilon^{-2}\: \frac{1}{R_\varepsilon^2} \\
T^r_r &= -\varepsilon^{-2}\: \frac{\Rqu^2}{3 R^2 (R_\varepsilon^2 +\Rqu^2)}
+ \varepsilon^{-4}\: \frac{\lambda}{3 R_\varepsilon^3}\:\frac{1}{\sqrt{R_\varepsilon^2+\Rqu^2}}\: w_3 \:.
\end{align*}
For small~$\varepsilon$, the last summand of~$T^r_r$ dominates, which in view
of~\eqref{w3sign} is positive.
\QED
The statement of Proposition~\ref{prpsec} rules out any bouncing solutions which satisfy
the strong energy condition. This also seems necessary for consistency, because otherwise
such bouncing solutions would violate the Hawking-Penrose singularity theorems
(see in particular~\cite[Theorem~2 in Section~8.2]{hawking+ellis}).
On the other hand, the singularity theorems do not rule that bouncing solutions might satisfy
the dominant or weak energy conditions. Proposition~\ref{prpeclimit} shows that
the methods of Section~\ref{secbconstruct} do {\em{not}} give rise to a procedure
to construct such bouncing solutions. However, it is still possible that such solutions do exist
in a certain parameter range where~$\varepsilon$ is bounded away from zero.
Indeed, by choosing~$w_3$ at the bouncing time positive but sufficiently small,
one can arrange in view of~\eqref{TB1} that the dominant and weak energy conditions
are satisfied near the bouncing point.
A numerical study showed that for such initial conditions, a typical solution does not
enter the classical dust region, but instead the scale function has a maximum after a short time and
then decreases again, leading to a singularity.
The question whether there are special initial conditions where the solution does enter the
classical dust region seems a hard global problem which we cannot enter here.

\section{Existence of Time-Periodic Solutions}\label{existperiodic}
For the construction of time-periodic solutions we shall make use of the following
symmetry argument.
\begin{Lemma} \label{lemmasymm}
Let~$(R_\eps, \vec{w}_\eps)$ be a solution of the rescaled Bloch
equations~\eqref{rescbloch} defined on a time interval~$[T_0,T_1]$.
Assume that for some time~$T$ in this interval,
\beq \label{initc}
\dot{R}_\varepsilon(T)=0 \qquad \text{and} \qquad
(w_\varepsilon)_2(T) = 0 \:.
\eeq
Then for all~$t$ with~$T \pm t \in [T_0, T_1]$,
\[ R_\varepsilon(T+t) = R_\varepsilon(T-t) \qquad \text{and} \qquad
\vec{w}_\varepsilon(T+t) = P \,\vec{w}_\varepsilon(T-t) \:, \]
where the operator~$P$ flips the sign of the second component,
\[ P \begin{pmatrix} w_1 \\ w_2 \\ w_3 \end{pmatrix}
= \begin{pmatrix} w_1 \\ -w_2 \\ w_3 \end{pmatrix} . \]
\end{Lemma}
\begin{proof} One easily verifies that both~$(R_\varepsilon(T+t), \vec{w}(T+t))$
and~$(R_\varepsilon(T-t), P \vec{w}(T-t))$ are solutions of the rescaled Bloch
equations~\eqref{rescbloch}
(note that the corresponding $R$-equations are the same, whereas in the~$\vec{w}$-equations
the second term in the brackets on the right has opposite signs;
the signs of the components of~$\vec{w}$ are chosen such the $\vec{w}$-equations hold in both cases).
According to~\eqref{initc}, these solutions have the same initial values at~$t=0$.
Therefore, we would like to apply the uniqueness theorem for ODEs to conclude that the solutions coincide.
However, this is not possible directly because the $R$-equation in~\eqref{rescbloch} is singular
if~$\dot{R}=0$. In order to bypass this problem, we consider the corresponding
vectors~$(R, \dot{R}, \vec{w})$ as solutions of the system of ODEs
consisting of the $\vec{w}$-equation in~\eqref{rescbloch},
the equation~$\frac{d}{dt} R = \dot{R}$ and the second order equation~\eqref{ein2} (also
rescaled by~\eqref{scal1}). Then the singularity at~$\dot{R}=0$ has disappeared, making
it possible to apply standard uniqueness results.
\end{proof}

\begin{proof}[Proof of Theorem~\ref{Thmperiodic}]
For given~$\lambda$ and~$R_I$, we let~$(R_\varepsilon, \vec{w}_\varepsilon)$
be a family of solutions of the rescaled Bloch equations~\eqref{rescbloch} with initial conditions
\[ R_\varepsilon(0)=R_I\:,\qquad \dot{R}_\varepsilon(0)=0 
\qquad \text{and} \qquad (\vec{w}_\varepsilon)_2(0)=0\:, \quad (\vec{w}_\varepsilon)_3(0)>0\:. \]
Following the constructions of Section~\ref{secbconstruct}, we know that
for every sufficiently small~$\varepsilon>0$, we have a bouncing solution
in the sense of Lemma~\ref{lemmasummary}.
Moreover, according to Lemma~\ref{lemmasymm} we know that
\beq \label{Rwsymm}
R_\varepsilon(-t) = R_\varepsilon(t) \qquad \text{and} \qquad
\vec{w}_\varepsilon(-t) = P\, \vec{w}_\varepsilon(t) \:.
\eeq
Hence there is~$T_{\max}>0$ where the scale function has a maximum.
If~$(w_\varepsilon)_2(T)$ vanishes, we see from~\eqref{Rwsymm} that our solution~$(R_\varepsilon,
\vec{w}_\varepsilon)$ has the same values at~$t=\pm T_{\max}$. Hence the uniqueness theorem for
ODEs yields that the solution can be extended to all times and is $2 T_{\max}$-periodic.
Thus it remains to show that for any~$R_I$, we can choose~$\varepsilon>0$ such
that~$(w_\varepsilon)_2(T)=0$.

Parametrizing~$\vec{w}_\varepsilon$ similar to~\eqref{phase1init}
by~$\vec{w}_\varepsilon = ((w_\varepsilon)_1, \rho_\varepsilon \cos \phi_\varepsilon,
\rho_\varepsilon \sin \phi_\varepsilon)$, this condition
is equivalent to demanding that~$\phi_\varepsilon(T_{\max}(\varepsilon)) \in \pi/2 + \pi \Z$.
In order to arrange this condition, we recall that $R_\eps(t)$ is continuous in $\eps$ for
each fixed $t$. Since $R_\eps$ is concave around its maximum value
for $\eps$ small enough, we see that $R_{\rm max}(\eps)$ is
continuous in~$\eps$, hence so is $T_{\rm max}(\eps)$. As a consequence, 
$\vec{w}_\varepsilon(T_{\max})$ and thus
also~$\phi_\varepsilon(T_{\rm max}(\eps))$ is continuous in~$\eps$.
Next, combining the bounds~\eqref{Tmaxest} and~\eqref{w1bound2}, one
sees from the $\vec{w}$-equation that~$\phi_\varepsilon(T_{\rm max}(\eps))$ tends to infinity
as~$\varepsilon \searrow 0$. Thus from the intermediate value theorem,
there is~$\varepsilon$ such that~$\phi_\varepsilon(T_{\rm max}(\eps))
\in  \pi/2 + \pi \Z$.
\end{proof} \noindent
Note that the last proof even shows that for every fixed~$R_I$,
the admissible~$\varepsilon$ form an infinite series~$(\varepsilon_n)_{n \in \N}$ with~$\lim_{n \rightarrow
\infty} \varepsilon_n = 0$.
We also remark that additional time-periodic solutions could be constructed by
choosing~$(w_\varepsilon)_2(0) \neq 0$ and adjusting~$R_I$ and~$\varepsilon$ so as to arrange
that~$(w_\varepsilon)_2(T_{\min}) = 0 = (w_\varepsilon)_2(T_{\max})$.
Then in general~$R_\varepsilon(T_{\min}) \neq R_\varepsilon(T_{\max})$, but 
by applying Lemma~\ref{lemmasymm} one sees that the
solution is still time-periodic with the doubled period~$2 (T_{\max} - T_{\min})$.

\section{Solutions in an Open and Flat Universe} \label{sec6}
\begin{figure}
\includegraphics[width=6.5cm]{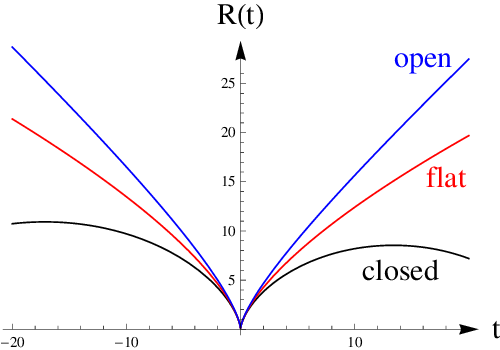} $\qquad$
\includegraphics[width=6.5cm]{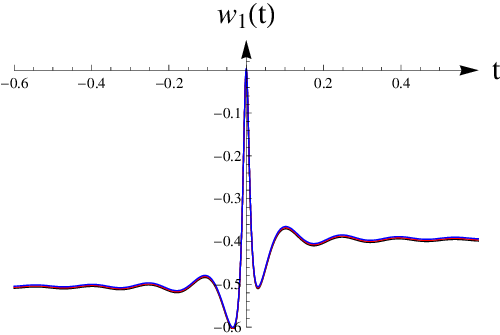}
\caption{Example of bouncing solutions in a closed, flat and open universe
for~$m=21.5$, $\lambda=3/2$, $\Rmax = 0.1$, $\Rtilt \approx 0.0698$.}
 \label{figopen}
\end{figure}

So far, we only analyzed the closed case. 
We now outline which of our methods and results carry over to the
the open and flat cases.
We first point out that the parameter~$k$ which distinguishes the different cases
only enters the $R$-equation in~\eqref{bloch} via the summand~$+k$.
In the closed case, the summand~$+1$ is essential for the universe to form an equator
in the classical dust limit. Thus in the open and flat cases, there is no mechanism which
contracts the universe after entering the classical era.
However, in our analysis of the quantum era, the summand $+k$ in the $R$-equation
was never used. Indeed, it even dropped out of the equations in the microscopic
limit~\eqref{blochlimit}. In particular, our construction of bouncing solutions
applies just as well in the open and flat cases. The only difference is that after entering
the classical era, the universe will keep expanding forever. This is illustrated in Figure~\ref{figopen}
in a typical example.
Clearly, no time-periodic bouncing solutions exist.

\begin{figure}
\includegraphics[width=6.5cm]{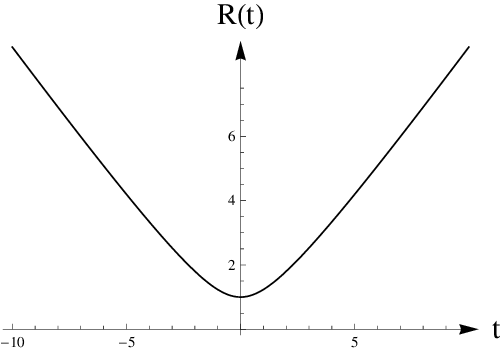} $\qquad$
\includegraphics[width=6.5cm]{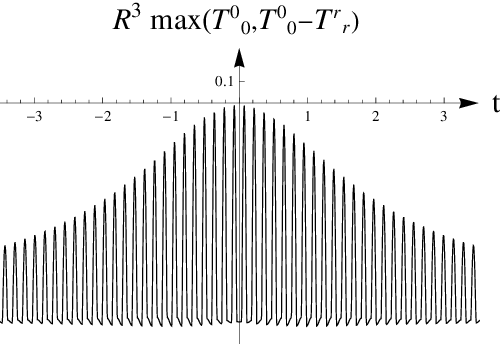}
\caption{Example of a bouncing solution in the open case with negative energy density
for~$m=21.5$, $\lambda=3/2$, $\Rmax = 1$, $\Rtilt \approx 0.0698$.
For this solution, $w_1 \approx 0.046$ is almost constant.}
\label{figneg}
\end{figure}

A specific difference in the open case is that the component~$w_1$ can be positive
(see the $R$-equation in~\eqref{bloch}). In view of~\eqref{TB1}, this gives rise
to bouncing solutions for which the energy conditions are violated even for large times
(see Figure~\ref{figneg} for a typical example).
These solutions correspond to the unphysical situation of a negative classical energy
density which drives an accelerated expansion of the universe.

\appendix
\section{The Dirac Equation and its Separation} \label{appA}
For the derivation of the corresponding Dirac operator
we follow the methods in~\cite{U22}, for an alternative derivation see~\cite{Villalba}.
For the Dirac matrices in Minkowski space we always choose the Dirac representation
\[  \gamma^0 = \left( \!\! \begin{array}{cc} \1 & 0 \\ 0 & -\1 \end{array} \!\! \right) , \qquad
\gamma^\alpha = \left( \!\!
            \begin{array}{cc}
            0 & \sigma^\alpha \cr
            -\sigma^\alpha & 0
            \end{array} \!\!
            \right) , \]
where~$\sigma^\alpha$, $\alpha=1,2,3$,  are the Pauli matrices,
\beq \label{PauliR3}
\sigma^1 = \begin{pmatrix} 0 & 1 \\ 1 & 0 \end{pmatrix}\:,\quad
\sigma^2 = \begin{pmatrix} 0 & -i \\ i & 0 \end{pmatrix}\:,\quad
\sigma^3 = \begin{pmatrix} 1 & 0 \\ 0 & -1 \end{pmatrix}\:.
\eeq
In curved space-time, the Dirac wave functions are four-component complex functions on the manifold,
which at every space-time point are endowed with an inner product~$\overline{\Psi} \Phi$ of
signature~$(2,2)$. For convenience, we represent this inner product just as in Minkowski space as~$\overline{\Psi}= \Psi^\dagger \gamma^0 \Psi$, thus~$\overline{\Psi}$ is the usual adjoint spinor.
The first step for the construction of the Dirac operator is to choose Dirac matrices~$G^j$ which are symmetric with respect to the inner product~$\overline{\Psi} \Phi$ and satisfy the anti-commutation relations
\begin{equation} \label{anticommute}
\{G^j,G^k\} \;\equiv\; G^j G^k + G^k G^j = 2 g^{jk} \1_4 \:.
\end{equation}
This can be accomplished by taking the following linear combinations of the Dirac matrices
of Minkowski space,
\beq \left. \begin{split}
 G^0 &= \gamma^0 \\
 G^r &= \frac{f(r)}{R(t)} \left( \cos\vartheta \ \gamma^3 +\sin\vartheta \cos\varphi \ \gamma^1 +
 \sin\vartheta \sin\varphi \ \gamma^2 \right) \\
G^\vartheta &=  \frac{1}{r R(t)} \left( -\sin\vartheta \ \gamma^3 +
 \cos\vartheta \cos\varphi \ \gamma^1 + \cos\vartheta \sin\varphi \ \gamma^2 \right) \\
G^\varphi &= \frac{1}{r R(t)\,\sin\vartheta} \left(-\sin\varphi \ \gamma^1 +
 \cos\varphi \ \gamma^2 \right) .
\end{split} \qquad \right\} \label{Gdef}
\eeq
The spin connection~$D$ can then be written as
\[ D_j = \partial_j -iE_j  \:, \]
where~$E_j$, the so-called spin coefficients, are given in terms of the Dirac matrices
and their first derivatives. Thus the Dirac operator becomes
\[ \D = i G^j D_j = i G^j \partial_j + (G^j E_j) \:. \]
For an orthogonal metric, the combination~$G^j E_j$ takes the simple form
(for details see~\cite[Lemma~2.1]{moritz})
\[ G^j E_j = \frac{i}{2 \sqrt{|g|}}\:
\partial_j(\sqrt{|g|} G^j) \quad \text{with} \quad g=\det g_{ij} \:, \]
making it unnecessary to compute the spin connection coefficients or even the
Christoffel symbols. A short calculation gives
\beq \label{dirac}
\D = i \gamma^0 \left( \partial_t + \frac {3 \dot R(t)}{2R(t)}\right) + \frac 1{R(t)} \left(
\begin{matrix}
0 & \D_{\mathcal{H}} \\ -\D_{\mathcal{H}} & 0 \\
\end{matrix} \right) ,
\eeq
where the purely spatial operator~${\mathcal{D}}_{\mathcal{H}}$ is given by
\begin{equation} \label{intrinsic}
{\mathcal{D}}_{\mathcal{H}} = i\sigma^r \left(
\partial_r +  \frac{f-1}{rf} \right) +i\sigma^\vartheta
\partial_\vartheta + i \sigma^\varphi \partial_\varphi \:,
\end{equation}
and the matrices~$\sigma^\alpha$, $\alpha \in \{\chi, \vartheta, \varphi\}$, are the following
linear combinations of the Pauli matrices,
\beq \label{Pauli}
\left. \begin{split}
 \sigma^r &:= f(r) \left( \cos\vartheta \ \sigma^3 +\sin\vartheta \cos\varphi \ \sigma^1 +
 \sin\vartheta \sin\varphi \ \sigma^2 \right) \\
\sigma^\vartheta &:=  \frac{1}{r} \left( -\sin\vartheta \ \sigma^3 +
 \cos\vartheta \cos\varphi \ \sigma^1 + \cos\vartheta \sin\varphi \ \sigma^2 \right) \\
\sigma^\varphi &:= \frac1{r\,\sin\vartheta} \left(-\sin\varphi \ \sigma^1 +
 \cos\varphi \ \sigma^2 \right) .
 \end{split} \qquad \right\}
\eeq

In order to separate the Dirac equation
\beq \label{Dirac}
(\D - m) \Psi = 0\:,
\eeq
we consider an eigenspinor~$\psi_\lambda$ of the spatial Dirac operator,
\[ {\mathcal{D}}_{\mathcal{H}} \,\psi_\lambda = \lambda \psi_\lambda \]
and employ the ansatz
\begin{equation} \label{ansatz}
\Psi = R(t)^{-\frac{3}{2}}
\left( \begin{matrix} \alpha(t) \:\psi_\lambda(r,\vartheta,\varphi) \\
\beta(t) \:\psi_\lambda(r,\vartheta,\varphi) \end{matrix} \right) ,
\end{equation}
we obtain the coupled system of ODEs~\eqref{DiracODE}
for the complex-valued functions~$\alpha$ and~$\beta$.
Using that the matrix in~\eqref{DiracODE} is Hermitian, one easily
verifies that
\[ \frac{d}{dt} \left( |\alpha|^2 + |\beta|^2 \right) = 0\:. \]
Thus we can normalize the spinors according to~\eqref{normalize}.

\section{The Energy-Momentum Tensor} \label{appB}
We begin in with the simpler closed case (the flat and open cases will be treated
in the paragraph following~\eqref{Trconvent} below). Then the operator~$\D_{\mathcal{H}} = \D_{S^3}$
with domain of definition~$C^\infty(S_3)^2)$
is an essentially self-adjoint operator on~$L^2(S^3)^2$. It has a purely discrete spectrum
with eigenvalues~$\lambda= \pm \frac{3}{2}, \pm \frac{5}{2}, \ldots$ and corresponding
eigenspaces of dimension~$N=\lambda^2 - \frac{1}{4}$ (see for example~\cite[Appendix~A]{moritz}).
For a given eigenvalue~$\lambda$ we let~$\psi_\lambda^{(1)}, \ldots, \psi_\lambda^{(N)}$ be an
orthonormal basis of eigenvectors. Using the ansatz~\eqref{ansatz} (for fixed functions~$\alpha$
and~$\beta$), we introduce corresponding one-particle wave functions~$\Psi_1, \ldots, \Psi_N$.
Taking their wedge product,
\[ \Psi := \Psi_1 \wedge \cdots \wedge \Psi_N \:, \]
we obtain a state of the fermionic Fock space. This state is a homogeneous and isotropic Hartree-Fock state.

The energy-momentum tensor is introduced as the following expectation value,
\[ T_{jk}(x) = \frac{1}{2} \: \text{Re} \,\big\langle \overline{\Psi(x)}
(i G_j D_k + i G_k D_j) \Psi(x) \big\rangle_{\mathscr{F}}\:, \]
taken with respect to the scalar product on the Fock space associated
to the probability integral, and where~$\Psi(x)$ and~$\overline{\Psi(x)}$
are the corresponding field operators acting on the the Fock space.
Using the orthonormality of one-particle wave functions~$\Psi_i$, the above expectation
value reduces to the sum of the one-particle expectations,
\[ T_{jk} = \frac{1}{2} \: \text{Re} \sum_{a=1}^N \overline{\Psi_a}
\left( i G_j D_k + i G_k D_j \right) \Psi_a \:. \]

In order to bring the energy-momentum tensor into a more explicit form, it is convenient
to introduce to introduce the object
\beq \label{Pform}
P(t,\vec{x};t',\vec{x}') = \sum_{a} \Psi_a(t,\vec{x}) \overline{\Psi(t',\vec{x}')}\:.
\eeq
Denoting the spectral projectors of~$\D_{S^3}$ by~$E_\lambda$ and
keeping in mind that the adjoint spinor in~\eqref{Pform} involves multiplying by the
matrix~$\gamma^0$, we can write~\eqref{Pform} as
\beq \label{Pansatz}
P(t,\vec{x};t',\vec{x}') = \left( R(t)\, R(t') \right)^{-\frac{3}{2}}
 E_\lambda(\vec{x},\vec{x}') \otimes \left( \begin{matrix} \alpha(t) \overline{\alpha(t')}
& -\alpha(t) \overline{\beta(t')} \\
\beta(t) \overline{\alpha(t')} & -\beta(t) \overline{\beta(t')}
\end{matrix} \right) .
\eeq
The energy-momentum tensor can be expressed in terms of~$P$ by
\beq \label{Tgen}
T_{jk}(t,\vec{x}) = \frac{1}{2}\: \Tr_{\C^4} \Big\{ (i G_j D_k + i G_k D_j)
P(t,\vec{x}; t', \vec{x}') \Big\}
\big|_{t'=t, \vec{x}'=\vec{x}} \:,
\eeq
where the trace is taken over $4 \times 4$-matrices.
Now we could proceed by computing the spin connection and substituting the resulting
formulas into~\eqref{Tgen}. The following shorter method avoids the computation of the
spin coefficients: Collecting the time-dependent part
in~\eqref{dirac}, we see right away that~$D_t = \partial_t -3 \dot{R}/(2R)$.
Hence
\[ T^0_0 = \Tr_{\C^4} \Big\{  i G^0 \Big( \partial_t + \frac{3 \dot{R}(t)}{2 R(t)} \Big) P(t,\vec{x};t',\vec{x}')
\Big\} \Big|_{t'=t, \vec{x}'=x} \:, \]
and employing the ansatz~\eqref{Pansatz}, we obtain
\beq \label{Ttt}
T^0_0 = R^{-3} \Big[ m \Big(|\alpha|^2-|\beta|^2 \Big)
- \frac{2 \lambda}{R}\: \text{Re}(\alpha \overline{\beta}) \Big]\,
\Tr_{\C^2} E_\lambda(\vec{x}, \vec{x})\:.
\eeq
Next it is easy to compute the trace of the energy-momentum tensor.
Namely, using that~$\D = i G^j D_j$, we get from~\eqref{Tgen}
\begin{align}
T^k_k &= \Tr_{\C^4} \left\{ \D P(t,\vec{x}; t', \vec{x}') \right\}
\big|_{t'=t, \vec{x}'=\vec{x}}
= m \Tr_{\C^4} \left\{ P(t,\vec{x}; t', \vec{x}') \right\}
\big|_{t'=t, \vec{x}'=\vec{x}} \nonumber \\
&= R^{-3} \:m \Big(|\alpha|^2-|\beta|^2 \Big) \Tr_{\C^2} E_\lambda(\vec{x}, \vec{x})\:. \label{trT}
\end{align}
The quantity~$\Tr_{\C^2} E_\lambda(\vec{x}, \vec{x})$ appearing in~\eqref{Ttt} and~\eqref{trT} is
a constant. It is computed by
\[ \Tr_{\C^2} E_\lambda(\vec{x}, \vec{x}) = \frac{1}{\mu(S^3)} \int_{S^3}
\Tr_{\C^2} E_\lambda(\vec{x}, \vec{x})\: d\mu(\vec{x}) =  \frac{1}{\mu(S^3)}\: \tr(
E_\lambda) = \frac{1}{\mu(S^3)} \left( \lambda^2 - \frac{1}{4} \right) . \]
In view of our normalization convention~\eqref{normalize} for~$\alpha$ and~$\beta$,
it seems most convenient to use the convention~$\mu(S^3) = 1$ and to set
\beq \label{Trconvent} \Tr_{\C^2} E_\lambda(\vec{x}, \vec{x})=1\:.
\eeq
Knowing~\eqref{Ttt} and~\eqref{trT} and using the convention~\eqref{Trconvent},
we can determine all components of the energy-momentum
tensor by a symmetry consideration. First, spherical symmetry implies that~$T^0_r=T^0_\vartheta=T^0_\varphi=0$. Next, using that our system is homogeneous in space, we conclude
that~$T^r_\vartheta=T^r_\varphi=T^\vartheta_\varphi=0$. Thus the energy-momentum tensor
must be a diagonal matrix. Using once again that our system is homogeneous, we
even obtain that~$T^r_r = T^\vartheta_\vartheta = T^\varphi_\varphi$. Hence the
energy-momentum tensor has only two independent components, which are determined
from the two equations~\eqref{Ttt} and~\eqref{trT}. We thus obtain the equations~\eqref{Tform},
and all other components of the energy-momentum tensor vanish.

In the open and  flat cases, the situation is a bit more difficult because
the operator~$\D_{\mathcal{H}}$ has a continuous spectrum, making it necessary to
proceed as follows. After choosing a self-adjoint extension of~$\D_{\mathcal{H}}$, we can
apply the spectral theorem
\beq \label{spectral}
{\mathcal{D}}_{\mathcal{H}} = \int_{\sigma({\mathcal{D}}_{\mathcal{H}})}
 \lambda\, dE_\lambda\:,
\eeq
where~$dE_\lambda$ is the spectral measure. Using that the spectrum is absolutely continuous,
we can represent the spectral measure by an integral kernel, i.e.
\[ dE_\lambda = E_\lambda d\lambda \qquad \text{with} \qquad
(E_\lambda \psi)(\vec{x}) = \int E_\lambda(\vec{x},\vec{y})\, \psi(\vec{y})\: d\mu(\vec{y})\:. \]
This integral kernel is no longer a projector. It is an operator of infinite rank, corresponding to the fact that our system now involves an infinite number of particles. Also, the states in the image of~$E_\lambda$
are no longer normalized, but we must rely on the distributional normalization
\[ E_\lambda E_\mu = \delta(\lambda- \mu)\: E_\lambda\:. \]
Describing the fermions our our system by~$E_\lambda(\vec{x},\vec{y})$, all {\em{local}} quantitites (like the
energy-momentum tensor, the current, etc.) are well-defined. But the total energy, charge, etc.\ will clearly diverge because the spatial volume is infinite. With this in
mind, we can work with~$E_\lambda(\vec{x},\vec{y})$ just as if it were a projector
onto the fermionic states of our system.
Then the above considerations again apply, and for convenience we take~\eqref{Trconvent} as
a normalization convention. This again gives the equations~\eqref{Tform}.

\Thanks{{\em{Acknowledgments:}} We are grateful to Christophe Morris for helpful comments.
We would like to thank the Erwin Schr\"odinger Institute, Vienna, for its hospitality and support. We are
grateful to the Vielberth foundation, Regensburg, for generous support.}


\def\dbar{\leavevmode\hbox to 0pt{\hskip.2ex \accent"16\hss}d}
\providecommand{\bysame}{\leavevmode\hbox to3em{\hrulefill}\thinspace}
\providecommand{\MR}{\relax\ifhmode\unskip\space\fi MR }
\providecommand{\MRhref}[2]{%
  \href{http://www.ams.org/mathscinet-getitem?mr=#1}{#2}
}
\providecommand{\href}[2]{#2}

\end{document}